%% file: crystal-arxiv.tex
\def\shortonly#1{}

\def\nonblind#1{#1}
\def\onlyblind#1{}
\def\ext#1{#1}

\documentclass{article}
\usepackage{graphicx}
\usepackage{url}
\usepackage{amsfonts}
\usepackage{amsmath}
\usepackage{amsthm}
\usepackage{booktabs}
\usepackage{subcaption}
\usepackage{multirow}
\usepackage{newfloat}
\usepackage{listings}
\usepackage[utf8]{inputenc}
\usepackage[T1]{fontenc}
\usepackage{hyperref}
\usepackage[ruled]{algorithm2e}

\def\bbZ{\mathbb{Z}}
\def\bbR{\mathbb{R}}
\def\bbH{\mathbb{H}}

\begin{document}

\author {
    Eryk Kopczyński, Dorota Celińska-Kopczyńska  \\
    { \small Institute of Informatics, University of Warsaw}
}

\title{Navigating Higher Dimensional Spaces using Hyperbolic Geometry}


\maketitle

\begin{abstract}
 Higher-dimensional spaces are ubiquitous in applications of mathematics. 
  Yet, as we live in a three-dimensional space, visualizing, say, a four-dimensional
  space is challenging. We introduce a novel method of interactive visualization
  of higher-dimensional grids, based on hyperbolic geometry. In our approach, visualized objects are 
  adjacent on the screen if and only if they are in adjacent cells of the grid. Previous attempts
  do not show the whole higher-dimensional space at once, put close objects in distant parts of the screen,
  or map multiple locations to the same point on the screen; our solution lacks these disadvantages,
  making it applicable in data visualization, user interfaces, and game design.
\end{abstract}

\section{Introduction}

The world we live in is three-dimensional: the location of every
object can be described by three coordinates ($x$, $y$, $z$; or $x_1$,
$x_2$, $x_3$). However, mathematically, there is nothing special about
the number 3: we can study the $d$-dimensional space $\mathbb{R}^d$ 
for any number of dimensions $d$.

Such higher-dimensional spaces are ubiquitous when we are modelling
our world mathematically. For example, colors are defined by their
three components (Red, Green and Blue), which makes the space of
possible colors three-dimensional, usually represented in computer
displays as three numbers from 0 to 255, or $\{0..255\}^3$.
However, some people and animals see four basic colors; for them,
the space of colors would be four-dimensional. Alternatively, 
we could have transparency as the fourth component, or we could
model the space of the possible smells.

These higher-dimensional spaces can be easily represented
mathematically, and handled efficiently by computers. With 
very high number of dimensions we may hit the so-called ``Curse of
Dimensionality'' \cite{curse} -- the resources required by 
the algorithms may depend exponentially on the number of dimensions.
While that should not be an issue with four or five
dimensions, even $\mathbb{R}^4$ is very hard for
humans to visualize.

In fact, even three-dimensional space may be challenging. 
This is firstly because our sheets of paper and computer screens
are two-dimensional, and secondly because our brains are
only used to relatively simple three-dimensional situations.
Navigating a three-dimensional maze, which uses all three dimensions
equally (i.e., it is not just a sequence of two-dimensional mazes)
is difficult for us. Our buildings, as well as maps and puzzles in
our 3D video games, typically have a simple structure.
Perspective works intuitively for representing these simple 3D
structures. In data analysis and machine learning, dimensionality
reduction algorithms such as Kohonen's Self-Organizing Maps (SOM)
\cite{kohonen} are used to reduce the dimensionality of the data to (usually) two
dimensions, in order to make it conveniently presentable to humans.

A common method of displaying 4D objects is the
Schlegel diagram \cite{coxeter}, which is basically a perspective
projection of the 4D object in $\mathbb{R}^3$; that object is
then projected again onto a 2D computer screen. This method
works pretty well for highly regular structures, such as the six
regular solids in $\mathbb{R}^4$, but more complicated shapes are
hard to interpret. Another common method of imagining the 4D space
is {\it time slicing}: using time as the fourth dimension, for example, by displaying
the consecutive 3D slices using perspective projection. Instead of 
using time, we can have {\it slider slicing}, where the extra coordinate may be changed manually using a 
slider or another UI element \cite{4dtoys}. Again, this method works well in some
situations, but is not viable when we want to see the whole 4D object
at once. Yet another common method is {\it spatial slicing}, where we put the slices next to 
each other; for example, by placing four 4x4 grids next to each other,
we can easily play three-dimensional Tic-Tac-Toe \cite{tictac3} on a piece of paper.
This process can then be repeated to show more dimensions, as seen in the 4D Ultimate
Tic-Tac-Toe\cite{tictacultimate} or 4D or even 6D versions of the puzzle
video game Minesweeper. Spatial slicing lets the player see the whole board clearly, but it makes it hard to
tell which cells are close to each other, and it is only applicable 
if all the dimensions are small. Control is another issue -- it is difficult
to find a good control scheme for moving in four dimensions, and rotation makes it even
harder (4D objects can be rotated in 6 axes).
It is hard to imagine an intuitive interface for that.

The volume of a ball of radius $r$ in $\mathbb{R}^d$ is proportional
to $r^d$. This makes the Euclidean space more complex in a greater
number of dimensions, and one may think that our problems with visualizing
higher dimensional spaces is a consequence of this complexity.
However, we also have non-Euclidean spaces, in particular the
$d$-dimensional hyperbolic space $\mathbb{H}^d$. Non-Euclidean spaces in two
dimensions include the sphere $\mathbb{S}^2$ and the hyperbolic plane
$\mathbb{H}^2$, which change our old Euclidean geometry in two opposite ways.
The sum of angles of a spherical $n$-gon is greater than $(n-2)180^\circ$
degrees, while the sum of angles of a hyperbolic $n$-gon is less; 
In $\mathbb{S}^2$ the straight lines converge, in $\mathbb{H}^2$ they diverge.
We do not see many examples of hyperbolic geometry in real life; 
real-world approximations of hyperbolic geometry include saddle shapes
and hyperbolic crocheting \cite{taimina}, however these work only for small
fragments of the hyperbolic plane. This is because while $\mathbb{S}^d$ is
a finite space, and the order of growth of $\mathbb{R}^d$ is $r^d$, the
hyperbolic space $\mathbb{H}^d$ grows exponentially: the volume of a ball
of radius $r$ is of order $e^{r(d-1)}$, and it simply cannot fit in our
Euclidean world, in any number of dimensions.

The structure of the hyperbolic space is similar to that of a tree, except
that it is continuous. Anyone who has tried to draw a tree on a piece of
paper will find out that at some point there is not enough space to draw
the further branches; however, it is straightforward to draw even large
trees in the hyperbolic plane. This property of the hyperbolic plane 
has found uses in the visualization \cite{Munzner,hyptree}. 
In applications such as
machine learning and social network analysis, it is common 
to embed the data into a metric space and predict their relationships
based on the distances between them. While the Euclidean space is the
most obvious choice, it turns out that 
when the data represented is of hierarchical nature, even
the two-dimensional hyperbolic geometry is much better than $\mathbb{R}^d$
\cite{papa,hsom}, even for very large values of $d$. 

Strikingly, while the hyperbolic plane is (in some sense) more complex 
and harder to understand mathematically than $\mathbb{R}^d$, it 
is surprisingly straightforward to navigate. People with no knowledge of
non-Euclidean geometry, when experimenting with a dynamic computer simulation
of it\cite{hyperrogue}, will have no problems with navigation, even though
they usually assume that the simulation shows a fish-eye projection of the
Euclidean plane, or a surface of a sphere. 

Thus, the hyperbolic plane is more complex, but easier to navigate, than
$\mathbb{R}^d$ for $d>2$. 
In this paper we present a method of navigating the higher dimensional
spaces on a computer intuitively; our method is based on hyperbolic geometry.

There are many applications
of higher-dimensional spaces where we are concerned only with a discrete grid.
As mentioned above, 
in data analysis, self-organized maps (SOM) usually map the data to two-dimensional
grids; our method lets us do analysis on higher-dimensional grids and still visualize them
using two dimensions. One important potential application is video games, where such discrete grids are common.
For a long time,
fantasy and science fiction writers have been exploring worlds behaving differently
to our own \cite{flatland}, and video games function as great tools to explore such worlds,
whether the difference is some kind of ``magic'' or something that could let the
player learn higher mathematics or physics intuitively, such as different values
of physical constants \cite{slowerspeedoflight} or different rules of geometry
\cite{hyperrogue}. While four-dimensional video games exist (Wikipedia currently lists
more than 30 of them \cite{wiki4}), none of the games
released so far have gained popularity, probably because of the navigation issues
introduced by the common methods of visualizing higher-dimensional geometries.
Our method could be used to create games giving the feel of these geometries without
the navigation issues. 

\paragraph{Structure of the paper}
In the next Section we recall the basics of hyperbolic geometry, and introduce our
method. The following Sections contain various potential applications of our
method, from user interface and data analysis to games. We conclude with a summary and
directions of further research.

\nonblind{\paragraph{Demo}
The demo is included in RogueViz \cite{rogueviz2021download} (RogueViz demos $\rightarrow$ visualizing higher-dimensional spaces).}

\section{Our construction}\label{ourcon}

\def\svgwidth{\linewidth}
\begin{figure}
\def\myfont#1#2{{\fontsize{100}{150}#2}}
\includegraphics[width=\svgwidth]{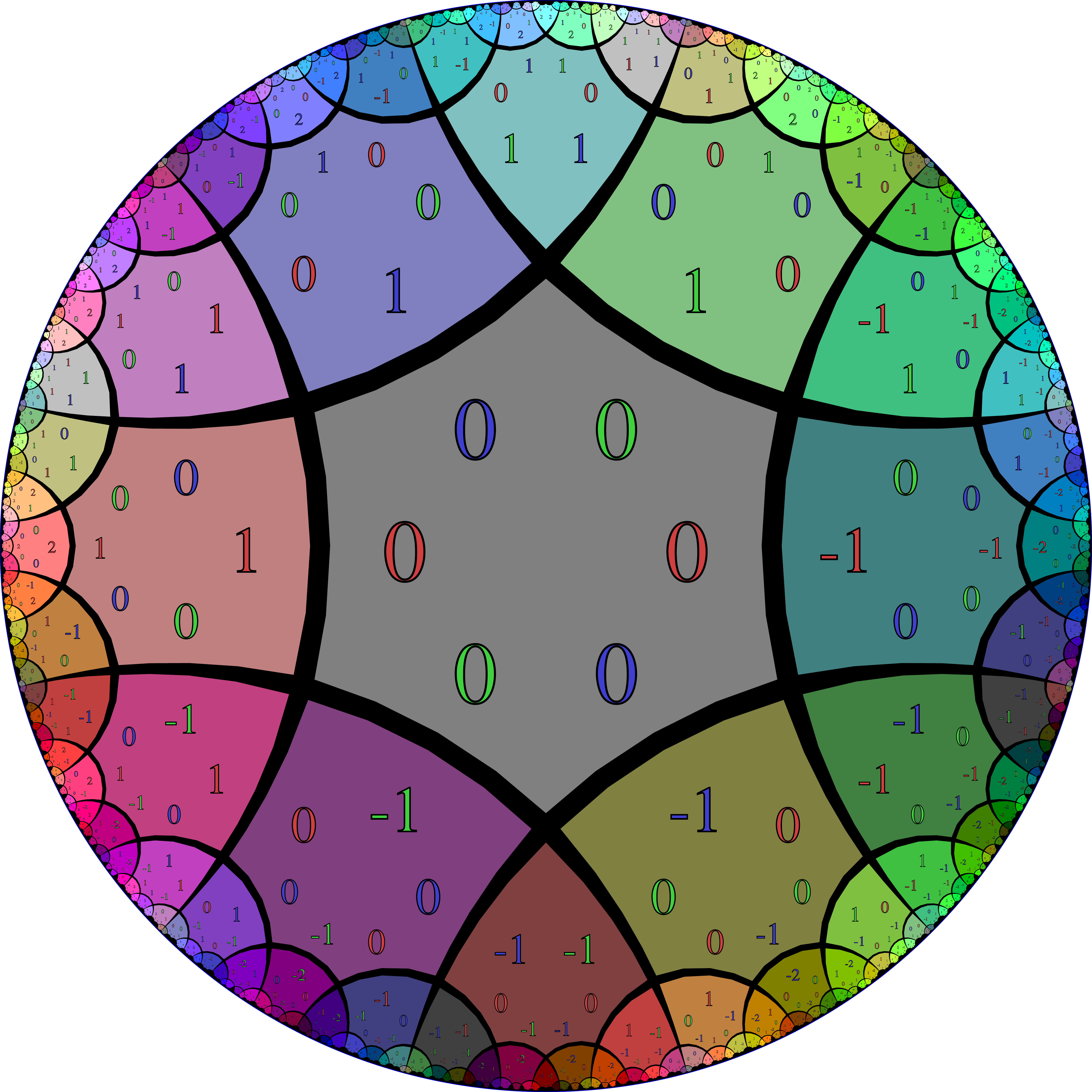}
\caption{Our construction in the Poincar\'e disk model. Three dimensions.\label{pmodel}}
\end{figure}

In this paper, as well as in our prototype implementation, the hyperbolic plane $\mathbb{H}^2$ will
be presented in the Poincar\'e disk model \cite{cannon}. The Poincar\'e disk model is the hyperbolic analog of the
stereographic projection of the sphere, and like all spherical and hyperbolic projections,
it introduces distortions. The whole hyperbolic plane is represented as the interior of
a disk, and straight lines are represented either by diameters of the disk, or by
circular arcs which intersect the boundary circle of the model at right angles.
Axial symmetry in a hyperbolic line $L$ is defined as the usual symmetry (when $L$ is
represented as a diameter) or inversion (when $L$ is represented as an arc). All the circular
arcs (except the outer circle) in Figure \ref{pmodel} are straight lines, and their pattern
is symmetric when any of these lines is taken as the axis of symmetry. 

We are using the Poincar\'e disk model because it is the most common method of
representing the hyperbolic plane. Because of its mathematical elegance it is usually
the first model shown in popular presentations of hyperbolic geometry, and furthermore it is
\emph{conformal}, which means that angles and small shapes are mapped faithfully.
This property gives aesthetically pleasing results, and often used by mathematical
artists including M.~C.~Escher \cite{coxe41}. It is worth to note that Poincar\'e disk
model is not the only possible projection -- by Riemann mapping theorem, conformal projections
exist to any simple connected subset of $\mathbb{R}^2$, and there are also non-conformal
projections too. In particular, the common mistake made by new programmers in hyperbolic
geometry is using the Poincar\'e disk model for internal calculations, while the Minkowski
hyperboloid model $\{(x,y,z) \in \mathbb{R}^3: x^2+y^2+1=z^2\}$ has better numerical properties,
and isometries in this model correspond directly to rotations of the sphere in the
$\{(x,y,z): x^2+y^2+z^2=1\}$ model. The hyperboloid model can be mapped to the Poincar\'e disk
model using the stereographic projection $(x,y,z) \mapsto (\frac{x}{1+z}, \frac{y}{1+z})$.

A regular tessellation can be described using its \emph{Schl\"afli symbol} \{p,q\}, where
$p,q\geq 3$. Symbol $\{p,q\}$ means
that the tessellation consists of regular $p$-gons, and $q$ of them meet in a single vertex.
There are three regular tessellations of the Euclidean plane: $\{4,4\}$ (square grid), $\{3,6\}$
(triangular grid) and $\{6,3\}$ (hexagonal grid); in each of them, we have $\frac{1}{p} + \frac{1}{q} = \frac{1}{2}$.
When $\frac{1}{p} + \frac{1}{q} > \frac{1}{2}$ we get a tessellation of the sphere (there
are five of them, corresponding to the five Platonic solids), and 
when $\frac{1}{p} + \frac{1}{q} < \frac{1}{2}$, we get a tessellation of the hyperbolic plane.
Our method of visualizing $\mathbb{Z}^d$ will be using the regular tessellation with Schl\"afli symbol $\{2d,4\}$
i.e., $2d$-gons such that four of them meet in every vertex. For $d=2$, we get the usual ``grid paper'' method of
visualizing $\mathbb{Z}^2$ on an Euclidean plane; for $d>2$, such as $d=3$ in Figure \ref{pmodel}, we will be using
a hyperbolic tessellation. Such hyperbolic tessellations exist for every $d>2$.

\begin{figure}
\includegraphics[width=\svgwidth]{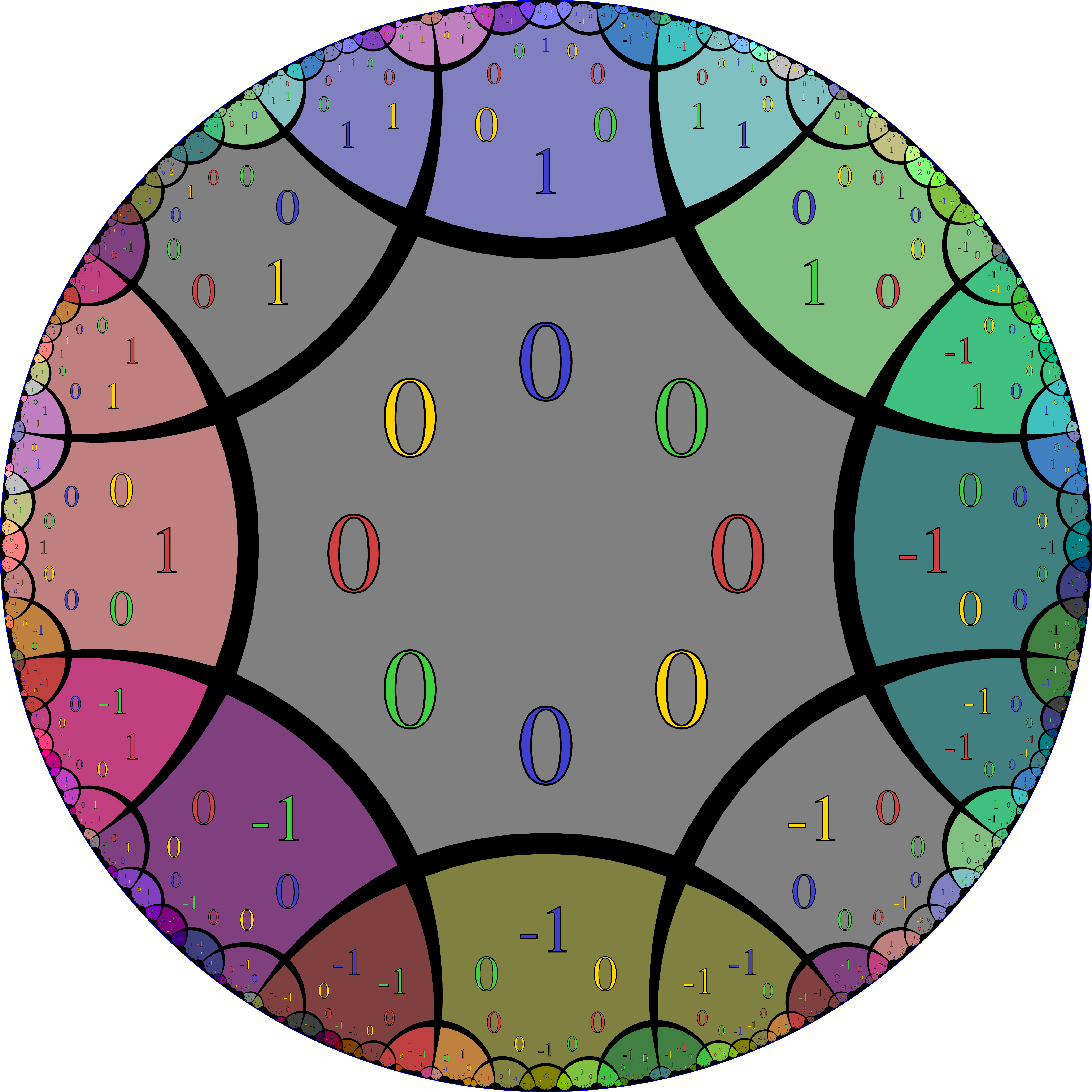}
\caption{Our construction in the Poincar\'e disk model. Four dimensions.\label{pmodel4}}
\end{figure}

Every $2d$-gonal tile $T$ of our tessellation will be assigned a point $p(T) \in \mathbb{Z}^d$. In Figure \ref{pmodel},
these points are listed using the colored numbers; each point has a Red, Green, and Blue coordinate. The central
$2d$-gon has all coordinates equal to 0. This choice is temporary -- using the translations of the hyperbolic plane,
the view can be smoothly shifted to be centered on any other point of focus \cite{hyptree}. In $\mathbb{Z}^d$, every point has $2d$ neighbors,
obtained by changing one of the coordinate values by 1. The $2d$ tiles adjacent to $T$ correspond to these $2d$ points adjacent
to $p(T)$: $T$ and $T'$ are adjacent iff $p(T)$ and $p(T')$ are adjacent. Additionally, if $T_1$ and $T_2$ are opposite
tiles adjacent to $T$, they also correspond to opposite neighbors of $p(T)$. In Figure \ref{pmodel}, the Red, Green, Blue
coordinates are not listed in a fixed order, but rather every coordinate is displayed close to the two opposite edges which
change this coordinate.

Other than that, the mapping of the tiles adjacent to $T$ to the points adjacent to $p(T)$ is arbitrary. Choose this mapping
for one tile determines the mapping to other tiles in the following way. Suppose we know the mapping $p$ for all the
tiles adjacent to $T_0$, and $T_1$ is one of these tiles. The rules given so far already determine the mapping for two tiles
adjacent to $T_1$ -- $T_0$ and the opposite tile. For the other tiles, let $L$ be the hyperbolic straight line between $T_1$ and
$T_2$. Let $T_2^+$ be one of the remaining $2d-2$ tiles adjacent to $T_2$, and $T_1^+$ be the tile adjacent to $T_1$ obtained
by symmetry in $L$. Then $p(T_2^+) = p(T_2) + p(T_1^+) - p(T_1)$, i.e., the direction in $\mathbb{Z}^d$ represented by 
going from $T_2$ to $T_2^+$ is the same as the direction from $T_1$ to $T_1^+$. In particular, when the tiles $T_1^+$ and $T_2^+$
are adjacent, and thus the four tiles $T_1$, $T_2$, $T_1^+$ and $T_2^+$ form a square, so do the mapped points $p(T_1)$,
$p(T_2)$, $p(T_1^+)$, $p(T_2^+)$.

The opposite rule guarantees that every straight line going through the centers of two adjacent tiles is mapped to a straight
line in $\mathbb{Z}^d$. The rules above also guarantee that every pair of two adjacent straight lines of this kind is mapped to
two adjacent straight lines in $\mathbb{Z}^d$, and wherever we cross the line between these two lines, it has the same effect
on the coordinates of the point we are in.

Figure \ref{pmodel4} is the same construction in four dimensions, with the fourth coordinate marked in golden.
In both cases, for each point $z \in \mathbb{Z}^d$, there are infinitely many tiles $T$ such that $p(T)=z$. This is not surprising,
since the number of tiles in $r$ steps from a fixed tile $T$ in a hyperbolic tessellation grows exponentially, while the number of points in $r$
from a fixed $z$ is of the order $r^d$.

\paragraph{Interpretation as a periodic structure}
Navigation in our construction has a natural interpretation in terms of a periodic $d$-dimensional surface. In this paragraph
we present this interpretation for $d=3$; similar interpretations exist in higher dimensions, but they are harder to visualize.

\begin{figure}
\includegraphics[width=\svgwidth]{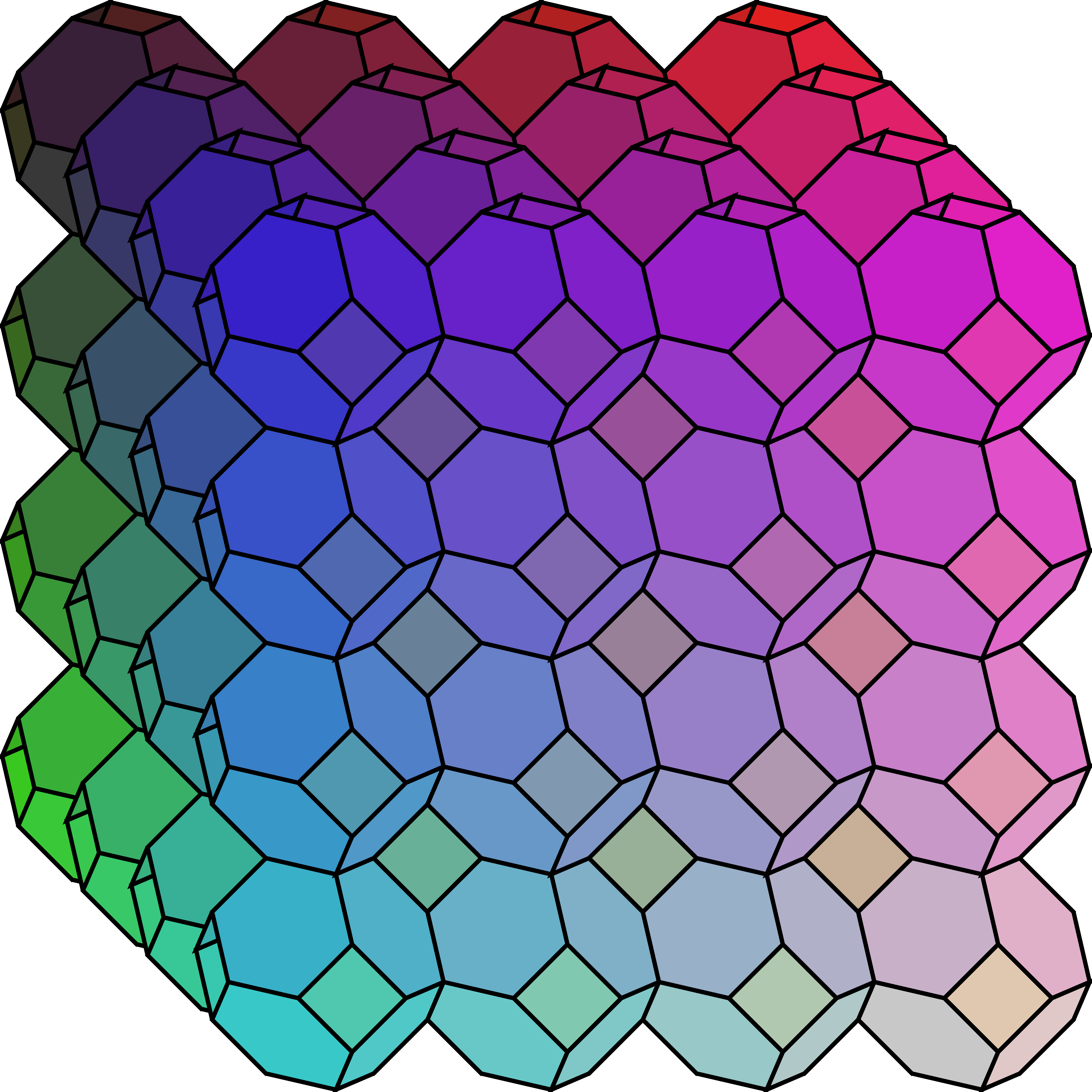}
\caption{The muoctahedron.\label{muocta}}
\end{figure}

Consider a unit cube $C$ in $\mathbb{R}^3$, centered in (0,0,0). The intersection of $C$ with the
plane $x+y+z=0$ is a regular hexagon $H$. Reflect that regular hexagon infinitely in each of the faces of the cube $C$. 
We get a infinite periodic structure (regular skew apeirohedron) known as the muoctahedron \cite{coxeter}, shown in
Figure \ref{muocta}.

Now, imagine we replace the original hexagon $H$ with a hyperbolic right-angled hexagon $H'$, in such a way that the four corner points
where $H$ met the edges of $C$ do not change, and $H'$ meets the faces of $C$ at right angles, so that the surface we obtain 
by mirror images is smooth. This way, we get a periodic surface $P$ of constant negative curvature, similar to the Schwarz minimal surface of type
P \cite{schwarz1890,schoen1970}.\footnote{In fact, minimal surfaces have zero mean curvature, while hyperbolic surfaces have constant negative Gaussian curvature.
By Hilbert's theorem, it is impossible to isometrically immerse a hyperbolic plane in $\mathbb{R}^3$. However, this construction is a very good
approximation. We could obtain hyperbolic geometry by changing the metric on Schwarz hexagon slightly, to match the metric on the
hyperbolic right-angled hexagon.}

Imagine that we are a (hyperbolic) flat creature living on $P$, and our physics are restricted to $P$; in particular, light rays do not leave $P$, but
rather travel along straight lines (geodesics) on $P$. As the light rays follow the surface $P$ and may bounce back, we will be able to see the same point,
or even images of ourselves, in multiple directions. (A simple analogy: if we were a flat creature living on a vertical cylinder of radius $r$, we would be able to
see a image of ourselves $k \cdot 2\pi r$ units from us in the horizontal direction, for every integer $k$.) Our visualization is thus exactly what 
such a hyperbolic flat creature would perceive: the hyperbolic plane from our visualization is the universal cover \cite{hatcher} of $P$.

Another intuition comes from the geometric group theory. Consider the $\{2d,\infty\}$ tessellation of the hyperbolic plane; there is a 
natural 1-1 correspondence between the tiles of this tessellation and the elements of the group $F^d$, the free group with $d$ generators.
The group $\mathbb{Z}^d$ is a quotient of $F^d$, hence we could also see the tiles as representing the elements of $\mathbb{Z}^d$ -- but
now there are multiple tiles corresponding to each group element ($ab$ and $ba$ are different in $F^d$, but the same in $\mathbb{Z}^d$).
Our construction improves on this by using some of the equivalences in $\mathbb{Z}^d$ to get a more compact representation.

\section{Three-dimensional hyperbolic space}
We can also visualize higher-dimensional Euclidean spaces using the three-dimensional hyperbolic space $\mathbb{H}^3$. 
Since we live in three-dimensional space ourselves, such visualizations are easy to interpret when projected to
two-dimensional displays, or when we use the Virtual Reality techniques to give an illusion of being inside a three-dimensional
hyperbolic space \cite{hyperbolicvr}. Our visualization is based on the same general ideas as the two-dimensional version:

\begin{itemize}
\item We use a regular honeycomb, i.e., a tessellation of our three-dimensional space where every cell is a regular solid.
\item Every cell $T$ has coordinates $p(T) \in \mathbb{Z}^d$. Its $2d$ faces correspond to $2d$ neighbors of $p(T)$.
\item If $T_1$ and $T_2$ are opposite cells adjacent to $T$, then they correspond to $p(T_1) + p(T_2) = 2p(T)$.
\item The dihedral angle between two faces is right (90 degrees).
\end{itemize}

See \cite{hhoney} for a reference about hyperbolic honeycombs. They are represented by Schl\"afli symbols $\{p,q,r\}$, where $\{p,q\}$
describes the shape of every cell of the honeycomb, while $\{q,r\}$ is the vertex figure. The regular solids in hyperbolic space
correspond to the regular solids in Euclidean space. The tetrahedron ($d=2$) does not work because it has no opposite faces;
the cube ($d=3$) yields the trivial Euclidean representation of $\mathbb{Z}^3$ based on the cubic honeycomb $\{4,3,4\}$.

For $d=4$ we get the honeycomb $\{3,4,4\}$. The 
\nonblind{video\footnote{\url{https://youtu.be/abRuuMeDwaY}}}
\onlyblind{included video} presents the visualizations of several four-dimensional structures based on
the $\{3,4,4\}$ honeycomb. This honeycomb has ideal vertices, i.e., we cannot reach the vertices of cells because
they are infinitely far away; the horospheres (surfaces orthogonal to all the edges ending in such an ideal vertex) 
are tessellated using the Euclidean square grid tessellation. Such a perspective is shown in the beginning of the video.
We fill some of the cells with solid $2d$-hedra based on their coordinates to visualize a four-dimensional structure (for example,
by filling all cells $T$  such that $p_1(T) \in \{1,-1\}$ we can visualize two parallel hyperplanes). 
Most basic four-dimensional structures yield nicely looking visualizations, reminding of zooms of fractals such as the Mandelbrot set
or three-dimensional fractals such as the Mandelbulb \cite{mandelbulb}, although the similarity appears to be superficial. While fractals
are more and more detailed as we zoom in, our constructions are constructed of flat surfaces, and fractal-like shapes in our visualizations
only appear in two-dimensional projections of our constructions, as projections of the ``infinite rays'', i.e., those which never hit any
cell. Our visualization includes:

\begin{enumerate}
\item (A) a 4x4x4x4 cage with a golden point in the center
\item (B) an one-dimensional tunnel (bright red)
\item (C) the 1-skeleton of the tessellation of $\mathbb{Z}^4$ with cubes of edge 2
\item (D) two-dimensional tunnel
\item (E) two hyperplanes in distance 2 (blue and green), i.e., three-dimensional tunnel
\item (F) two hyperplanes in distance 3 (cyan and green)
\item (G) two orthogonal hyperplanes (red and yellow)
\item (H) four quarterspaces (red, yellow, cyan, blue)
\item (I) diagonal tunnel in all coordinates except one (golden and silver)
\item (J) diagonal tunnel (purple and gray)
\end{enumerate}

While the number of honeycomb cells whose centers are in distance $d$ is exponential in $d$, such honeycombs can be rendered 
very efficiently in real time, using a fragment shader and raycasting. Suppose the camera is positioned in a point $h_0 \in T$ for some
cell $T$ of our honeycomb. For each pixel of the resulting image, we compute the direction $d$ of the ray in $\bbH^3$ corresponding to that pixel. For each face $F$
of $T$, we compute where the geodesic ray $(h_0,d)$ hits $F$. Take the face $F$ which is hit the first. If the cell $T_1$ on the other side
of the face $F$ is filled, color the pixel using the color of the filled cell, with some added texture depending on which point of $F$ was
hit and the total travelled distance. Otherwise, compute the new ray coordinates and direction relative to the cell $T_1$. 
Both computations (intersection of a ray and a hyperbolic plane, and the isometry moving $T_1$ to the center cell) can be done easily
in the Minkowski hyperboloid model.

Like the visualization of $\mathbb{Z}^3$ using $\{6,4\}$ can be interpreted using the surface of a muoctahedron (Figure \ref{muocta}), 
the visualization of $\bbZ^4$ using $\{3,4,4\}$ can be similarly interpreted analogously, using a four-dimensional version of the muoctahedron.
Tessellate $\bbR^4$ using cubes of edge 1 centered at the grid points. In the cube $C$ containing $(0,0,0,0)$, consider the set $S$ of 
all points $(x,y,z,w)$ such that $x+y+z+w=0$. The set $S \cap C$ is an octahedron, and this octahedron is a part of our three-dimensional
manifold that is mapped to a hyperbolic octahedron. Other octahedra are obtained by reflecting $S \cap C$ in the faces of the cubes.
(The video shows the time-sliced visualization of this structure.)

For $d=6$ we get the right-angled (order 4) dodecahedral honeycomb $\{5,3,4\}$. This honeycomb has material vertices, corresponding to 
the three-dimensional faces in $\bbZ^6$ where eight cells meet. The video also presents the visualizations of 
six-dimensional space using this honeycomb. These visualizations correspond to the four-dimensional structures 
listed above, with minor changes (in (E) the tunnel is four-dimensional, and in (C) 1-skeleton is shown). Because of the higher dimensionality,
it is more difficult to interpret these pictures. They 
For $d=10$ we would obtain the right-angled icosahedral
honeycomb $\{3,5,4\}$; this honeycomb has ultra-ideal vertices \cite{hhoney}. We have not attempted to visualize this case
because of the ultra-ideal vertices, and because the high dimensionality would make the visualization very hard to interpret.

\section{Applications}

\begin{figure}
\includegraphics[width=\linewidth]{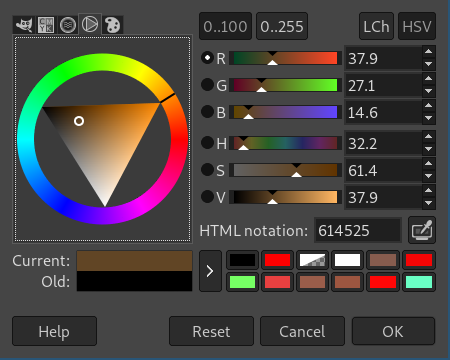}
\caption{Color picker in Gimp.\label{gimp}}
\end{figure}

\paragraph{Color picker}
As a very simple application of our method, we present a color picker. As mentioned in the Introduction, the space of possible colors is three-dimensional
(Red, Green, Blue in the RGB model, or Hue, Saturation, Value in the HSV model, which is better aligned with how humans perceive colors). The users of graphical design tools
such as Adobe Photoshop, GIMP, or Inkscape need a way to specify the color they want visually. The simplest method is to specify the three coordinates using three separate sliders.
However, this method has a disadvantage that it may be inconvenient to find the best color when this requires changing two values together (say, Saturation and Value). A better method,
available in all the tools above (Figure \ref{gimp}, is to specify only the Hue coordinate using a slider. This reduces the set of color choices to just two dimensions, which can be displayed as 
a triangle (Black, White, and fully saturated and bright color of the chosen hue), from which the specific color can be selected using a mouse. This method is basically {\it slider slicing}
applied to the three-dimensional space of colors.

The visualization in Figure \ref{pmodel} suggests that our method could be used as another solution to the color picking problem. Take $d=3$, and make every hexagon correspond to a color
based on its three coordinates; thus, increasing the Red coordinate corresponds to increasing the Red value by 1. All the hexagons are displayed in their corresponding colors, allowing
the user to see all the nearby colors on the single picture. One issue with this is that the Red, Green, Blue coordinates in the TrueColor model have values from 0 to 255; thus,
reaching the intended color from the initial color (say, (128, 128, 128)) would waste lots of time. To combat this issue, another slider is added, which controls by how much the coordinates
are changed with each step.

The same method can be used in any other situation where the user wants to specify the values of more than two coordinates, and the choices
can be examined visually (for example, trying out combinations of two colors, or picking a shape defined by three parameters). In data
visualization, line charts are perfect for visualizing functions of a single parameter ($f(x)=y$), and heat maps and surface charts
are good for visualizing functions of two parameters ($f(x_1,x_2)=y$); our method, either used by itself or in combination with the
aforementioned methods, could be used to visualize functions of more than two parameters.

\paragraph{Musical pitch spaces} One interesting potential application is a visualization of musical pitch spaces. Popular instruments
such as pianos arrange all the sounds by pitch in a one-dimensional way; however, when we consider how music is perceived by humans,
we naturally get a higher-dimensional model. Two sounds are consonant (i.e., they sound good together) when the ratio of their pitches is a simple fraction, such as
2:1 (octave), 3:2 (perfect fifth), 4:3 (perfect fourth), 5:4 or 7:5; in the equal temperament system of tuning used in most modern instruments,
these ratios are only approximated by powers of $\sqrt[12]{2}$. By composing the intervals listed above, we can get to any sound whose pitch ratio
to the original sound is $2^x 3^y 5^z 7^t:1$; thus, the space of sounds reachable using these intervals (called the 7-limit just intonation)
is four-dimensional.
Imagine that touching the octagon in Figure \ref{pmodel4} corresponding to point $(x,y,z,t)$ plays the sound whose pitch ratio to some base
sound is $(3:2)^x(4:3)^y(5:4)^z(7:5)^t$; this gives us a musical intrument, where octagons which are close play consonant sounds.
(The number of dimensions or ratios could be changed.) Technique similar to hyperbolic crocheting \cite{taimina} could be also used to bring
the hyperbolic plane, and thus also our instrument, to our real world.

\paragraph{Data analysis and visualization} Non-Euclidean geometries are recently gaining attention of the data scientists \cite{tda,tda_chazal}.
In particular, hyperbolic geometry has been proven useful in data visualization and modeling of scale-free networks \cite{hypgeo,papa}.
This usefulness comes from the exponential growth property of hyperbolic geometry, which makes it much more appropriate than Euclidean for modeling and visualizing hierarchical data. However, even if researchers tend to favor two-dimensional visualizations, we can imagine that our method should find application in visualizing of data dimension reduction techniques, such as Self-Organizing Maps, Multidimensional Scaling (MDS) or t-SNE. 


\section{Simple puzzles to understand the construction}
Games or puzzles are a great way to understand mathematical phenomena, especially those that are difficult to observe in three-dimensional
Euclidean space. In this section we present several simple games, which can be used to gain better intuitive understanding of
our visualization method.

\begin{figure}
\includegraphics[width=.5\svgwidth]{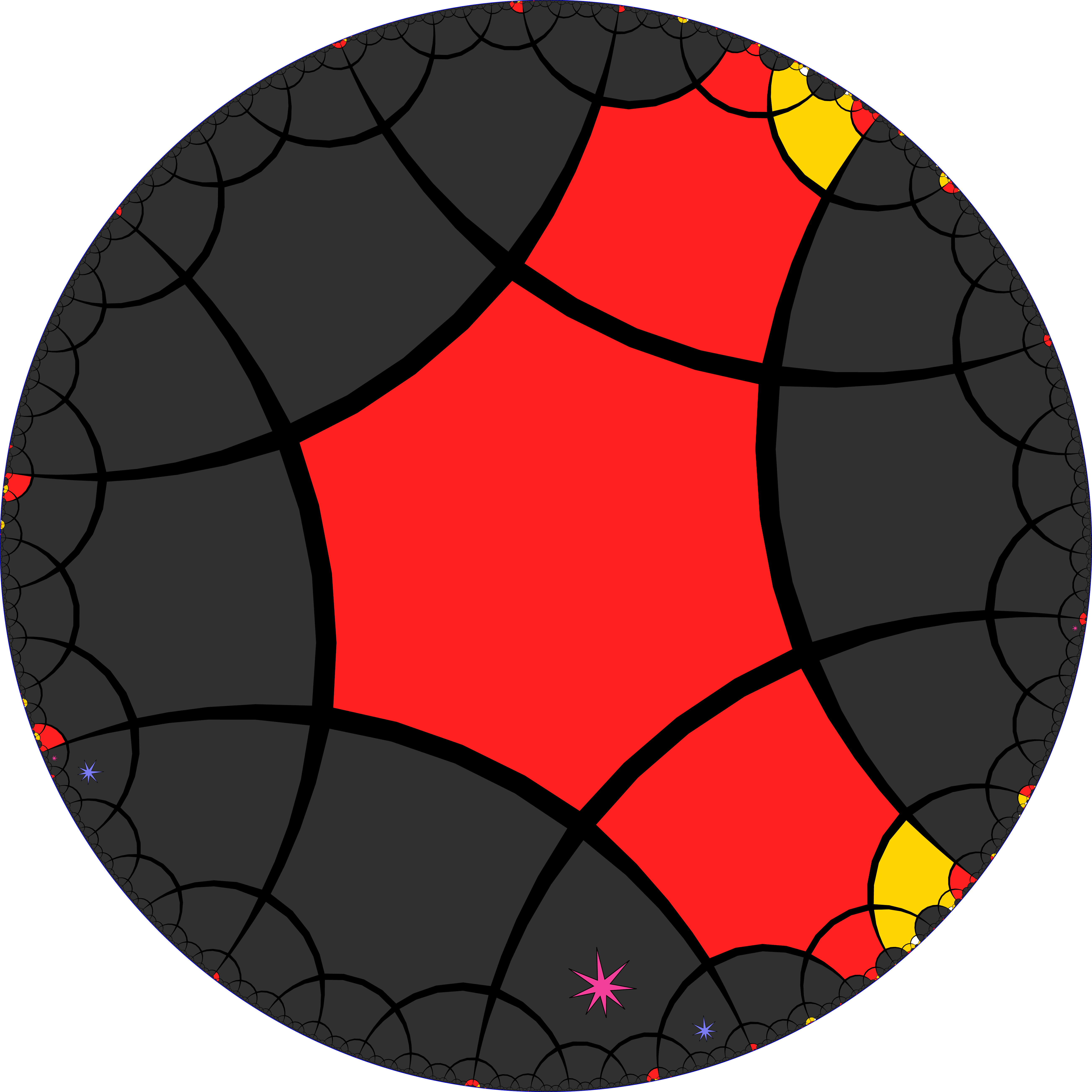} \hskip -1mm
\includegraphics[width=.5\svgwidth]{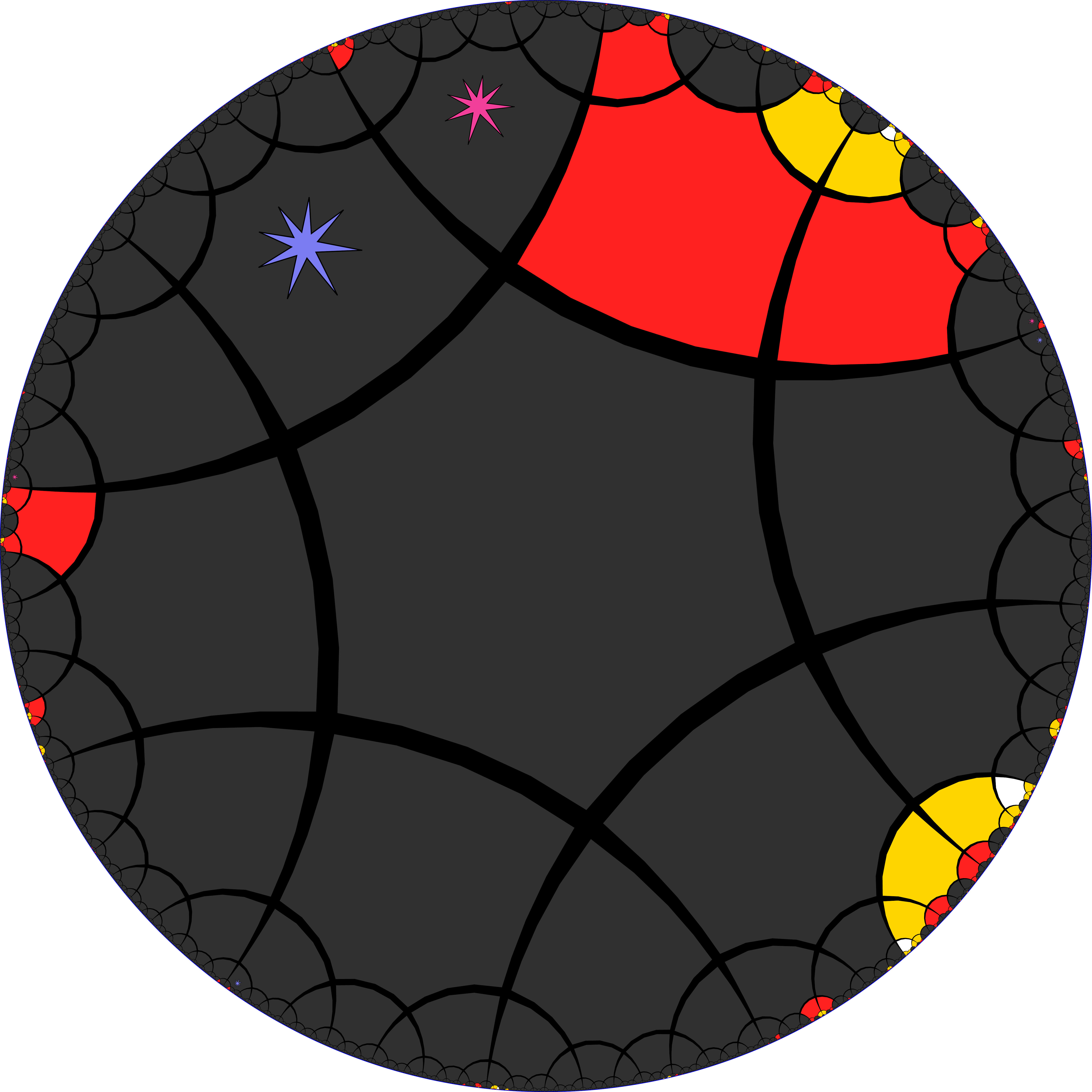}
\caption{Three-dimensional house, viewed from the corner and from the above.\label{house3}}
\end{figure}

\paragraph{Three-dimensional house} In {\it Flatland} \cite{flatland}, the Sphere proves its three-dimensional nature to the 
Square by using the third dimension to move Square's valuables out of his cupboard, which is locked, but only in two dimensions.
Similarly, a four-dimensional being would be able to see and manipulate things inside our locked three-dimensional houses.
This idea is used frequently in intuitive explanations of higher-dimensional spaces; the videos of the to-be-released 4D game
Miegakure \cite{miegakure} also show puzzles based on this idea. While this sounds like magic to the creatures living in the 
subspace, it is not for the higher-dimensional being performing the trick -- for them, it is the most natural thing, not a puzzle
or magic.

One of the puzzles in our prototype implementation asks the player to get inside a (small, $5\times 5\times 5$) three-dimensional cube,
placed in the four-dimensional space. As expected, this puzzle is very easy. Figure \ref{house3} shows the 
three-dimensional cube. Walls are red, inside is yellow, and the center is white. The left figure shows the cube viewed from its corner;
after we move a bit (above the house in the fourth dimension; stars in Figure \ref{house3} show the relation between the two views),
we get the right figure, and we see that the white center is easily reachable.

\paragraph{Find the center} 
Another puzzle asks the player to find the center of a simple four-dimensional shape, such as a hypercube (the set of points with all
coordinates with absolute values less than $r$; this generalizes 2D squares and 3D cubes) or an orthoplex (the set of points with the sum of
absolute values of coordinates less than $r$; this generalizes octahedra in 3D and 16-cells in 4D). Again, this simple puzzle 
allows the player to gain a feeling of the $d$-dimensional space. Finding the center should be obvious for a $d$-dimensional
being; however, our method is able to display only a small portion of the $d$-dimensional world on the screen, so to solve the 
puzzle, one would usually find the corner, and then move straight to the center via the diagonal, or go to the other corners
in order to determine the dimensions of the shape.

\begin{figure}
\includegraphics[width=\svgwidth]{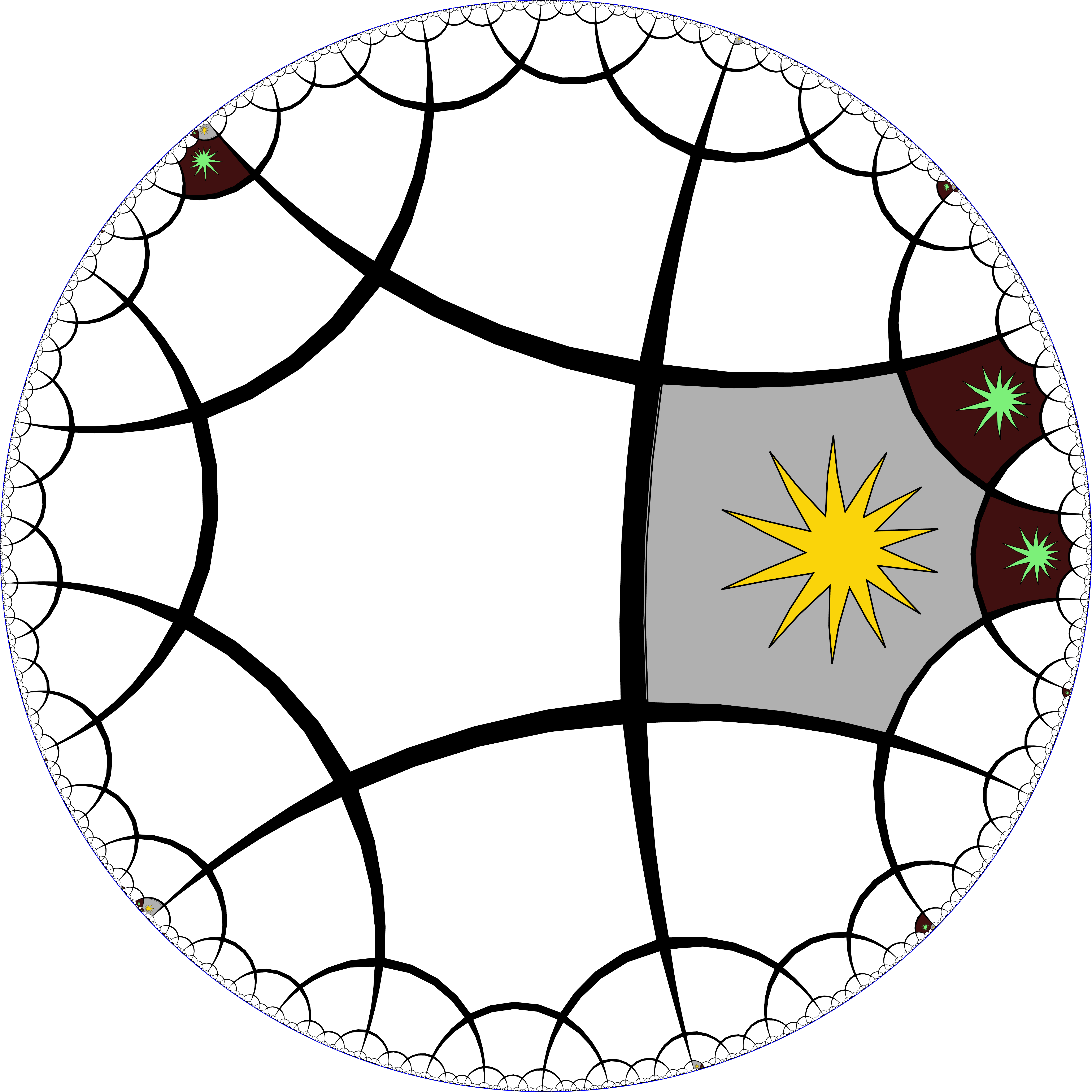}
\caption{Whatever direction the golden star moves, the two green stars will be able to move to two distinct locations
adjacent to it.\label{rlpic}}
\end{figure}

\paragraph{Basic roguelike} Roguelikes are a genre of single-player video games that share elements of puzzle, strategy,
and RPG genres \cite{craddock}. They are often compared to abstract strategy games such as Chess; however, in a roguelike,
the player is controlling a single powerful piece, while their opposition consists of multiple enemies, whose strength
tends to come from multiplicity and randomness rather than acting intelligently. In most roguelikes the player is modelled
with great detail as typical to the Role-Playing game genre, while the basic gameplay in HyperRogue \cite{hyperrogue} works
in a very simple way: the player and all the enemies alternate turns, in which they can move to an adjacent cell (if it is
not occupied), or attack a creature on the adjacent cell, destroying it; consequently, the player loses the game if they end
their turn next to an enemy. Enemies do not act intelligently, they simply move towards the player
character. In HyperRogue, even this very simple rule set leads to a somewhat interesting gameplay: in a hyperbolic tessellation,
it is easy to escape from, and eventually destroy, two monsters attacking at once, while in the usual Euclidean grids such as
the square grid $\{4,4\}$ or the hex grid $\{6,3\}$, this would be impossible.

While it is common for roguelikes to experiment with various structures of 2D maps, it is uncommon for them to experiment with
more than two dimensions, probably because the interesting consequences of playing in a 3D world tends to be outweighed by the
issues with clearly displaying and controlling the 3D environment. These issues can be solved with our method. One interesting
issue is shown on Figure \ref{rlpic}, where two enemies (green stars) are attacking our character (the golden star) at once.
It looks as if it should be possible to escape from two monsters
attacking at once (by moving in a direction where the monsters would have to both move to the player's previous position in
order to keep being adjacent); however, in practice this does not work, because while one monster moves to the player's position,
the other one will use its ``other copies'' to also keep their distance. This effect becomes obvious when considering the underlying
$n$-directional structure of the map. Thus, this situation can be used as a introduction to higher-dimensional geometry and our
visualization method.

\section{Applications to more complex games}

Our method can also be used to create more complex higher-dimensional games.
We give a list of properties that a game should have in order to benefit from our visualization.

\paragraph{Basics} The game should take place in the grid of hypercubes in three or more dimensions.

\paragraph{Gravity} In our world, gravity is a force which acts in a specific dimension and direction.
Gravity is also incorporated in the designs of many games. In our visualization method though, gravity
stops being intuitive, because the gravity direction seems to change between cells. This is not a
serious problem though, because three-dimensional games designed around gravity work very well with the usual
perspective methods. A four-dimensional game with gravity could be realized using a combination of our method and
perspective, as explained below.

\paragraph{Adjacency} The strength of our method is, that for every cell, the set of $2d$ cells directly adjacent
to it is represented in an intuitive way. However, games often consider cells adjacent if they share a vertex.
For example, in Minesweeper, every cell is adjacent to $3^2-1 = 8$, not $2\cdot 2=4$, other cells. In the natural
three-dimensional variant of Minesweeper, every cell will be adjacent to $3^3-1 = 26$ other cells, while our
method displays only 12 of them as corner-adjacent. This consists of the 6 cells where one coordinate changes,
6 out of 12 cells where two coordinate changes, and none of the 8 cells where all coordinates change. Of course, we
can ignore this and just make a three-dimensional version of Minesweeper with adjacency defined by the corner-adjacency
in our tiling, obtaining a game that, while having no obvious $n$-directional interpretation, may be interesting
in its own right.

\paragraph{Wide moving objects} Another thing that is hard to simulate using our method is large moving objects.
While pushing or rotation of such objects can be easily implemented, it will not work in an intuitive way. An archetypical
example of a game involving rotation of large objects is the well known Magic Cube, also known as Rubik's cube \cite{rubiks}.
More precisely, our method is not suitable for \emph{wide} objects. Things such as a snake
whose head pulls a long body together with it, or a plant that grows, are \emph{narrow}, and thus are not an issue.
This restriction seems to come from the hyperbolic plane, where wide moving objects are hard to work with, because 
in a ball of large radius $r$ moving alone a hyperbolic line $L$, points in distance $r$ from the line $L$ will move
much faster than the center of the ball \cite{hyperrogue} -- precisely, they will move $\sinh(r)$ times faster, where $\sinh$
is an exponentially growing function.
Interestingly, four dimensional variants of Rubik's cube can be visualized intuitively with a form of Schlegel diagram \cite{magic4},
and even seven-dimensional variants have been visualized using a combination of the Schlegel diagram and spatial slicing \cite{magic7}.
Another interesting perspective is visualizing the graph $R$ of configurations of the 3D magic cube. This graph has 
43 252 003 274 489 856 000 vertices and two configurations are adjacent if one can be reached from the other in a single move,
making it a degree 12 graph (each of the 6 faces can be rotated in two directions); it could be potentially visualized using
a method similar to our visualization of the 6-dimensional space, where moving in the $k$-th direction corresponds to rotating
the $k$-th face (since rotating the $k$-th face then rotating the $l$-th face usually has different effect than rotating first $l$ then $k$,
the details will have to be changed; however, $R$ is a group, and some visualization is possible according to the group theory intuitions
given in Section \ref{ourcon}).

\begin{figure}
\includegraphics[width=\svgwidth]{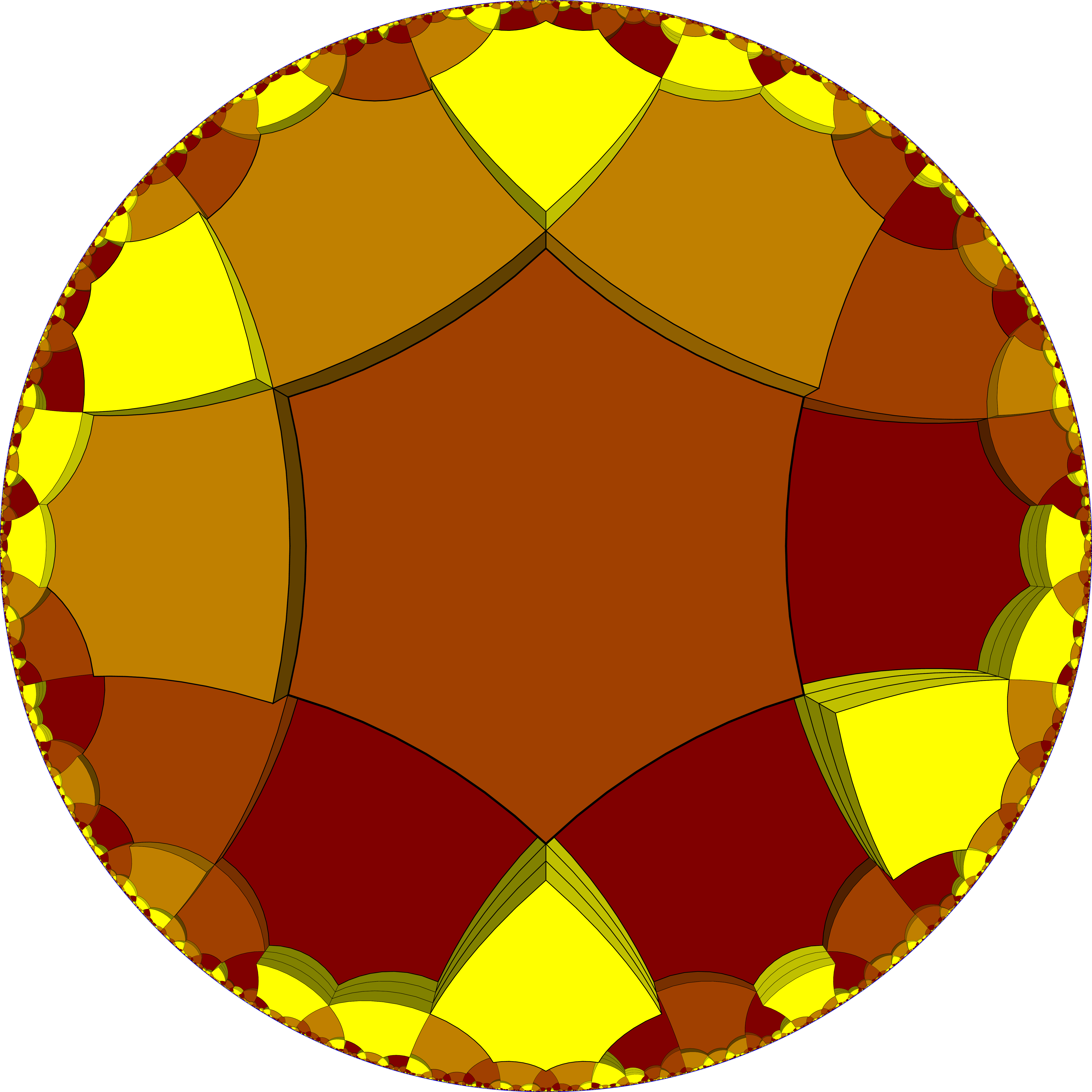}
\caption{Mockup of a 4D game with gravity.\label{mockup}}
\end{figure}

\paragraph{What games would work well?} An example of a game which would play well with out method is a higher-dimensional generalization
of Sokoban. Sokoban is a well-known block puzzle game played on a 2D grid. The player character moves in a warehouse
filled with walls and with some boxes placed. The player can move in four directions, they can also push a box in a straight
direction. To win the game, all the boxes must be moved to their designated destinations.
Since our method maps orthogonal straight lines
to hyperbolic straight lines, pushing of single-cell blocks works intuitively. We can make Sokoban 4D with gravity: 
while three dimensions will be mapped to the
plane using our method, the fourth (altitude) uses a form of perspective projection. Recall that the Poincar\'e disk model
is obtained from the hyperboloid model via the stereographic projection $(x,y,z) \mapsto (\frac{x}{w+z}, \frac{y}{w+z})$ for $w=1$. 
Changing the value of $w$ based on the altitude gives an intuitive looking picture (Figure \ref{mockup}).

\section{Comparison with previous methods}
Our method actually 
works for a discrete grid where all the coordinates can only be integers,
i.e., $\mathbb{Z}^d$ rather than $\mathbb{R}^d$. 
Since $\mathbb{H}^2$ grows faster than the Euclidean space, every point in
$\mathbb{Z}^d$ will be mapped to multiple points in $\mathbb{H}^2$, in 
a periodic way. (Topologically, our method maps $\mathbb{Z}^d$ to a quotient
space of the hyperbolic plane; we display the universal cover of this quotient
space, hence the periodic representation.)
Our approach advantages over the previous methods. The slicing methods do not show the whole space at once (time/slider slicing)
or show close points at distant locations (spatial slicing); the Schlegel
diagram method maps a whole $(d-2)$-dimensional subspace to a single point.
Our method maps points of $\mathbb{Z}^d$  to close points on the screen if and only if they are close in 
$\mathbb{Z}^d$ (as mentioned before, every point $x^i \in \mathbb{Z}^d$ will be 
mapped to many points $x^i_1, x^i_2, \ldots \in \mathbb{H}^2$; we assume that
we are measuring the distance between $x^1_i$ and $x^2_j$ which are the closest;
in other words, we measure distance in the quotient space).
\section{Conclusion}

We have presented our method of visualizing higher-dimensional Euclidean spaces using two-dimensional and three-dimensional
hyperbolic space. There are many possible directions of further research. One question is whether a similar method be used to visualize
higher-dimensional hyperbolic spaces (say, a honeycomb in $\mathbb{H}^3$) in $\mathbb{H}^2$. It is also possible that some details about
our method could be changed to reduce the distortions introduced by our method, which make it unintuitive for some applications (such as the games with gravity).

\nonblind{This work has been supported by the National Science Centre, Poland, grant UMO-2019//35/B/ST6/04456.}

\bibliographystyle{alpha}
\input{crystal-arxiv.bbl}

\end{document}

%% file: crystal-arxiv.bbl
\newcommand{\etalchar}[1]{$^{#1}$}